\documentclass[letterpaper, 10 pt, conference]{ieeeconf}  

\IEEEoverridecommandlockouts                              
\usepackage{amsmath, amssymb}

\usepackage{xcolor}
\usepackage{listings}
\usepackage{cite}
\usepackage{graphicx}
\usepackage{float}
\usepackage{mathtools}
\usepackage{empheq}
\usepackage{import}    
\usepackage{booktabs}
\usepackage{cleveref}

\title{\LARGE \bf
Risk-Averse Lander Site Selection under Altitude-Limited Information
}

\author{Vikas A. Patel, Mahdi Al-Husseini, Duncan Eddy, and Mykel J. Kochenderfer
\thanks{The authors are with Stanford University. Emails: 
        {\tt\small \{patelva, mah9, deddy, mykel\}@stanford.edu}}%
}

\begin{document}

\maketitle
\thispagestyle{empty}
\pagestyle{empty}

\begin{abstract}
In aerospace systems, powered descent requires efficiently selecting a landing site while fine-scale hazards remain unresolvable until low altitude. This process presents a decision challenge since the actor must select a site and make corresponding actions before all information is known. To successfully solve this problem, an agent must reason over potential risks and make corrections as new observations are made. We introduce a lightweight model of altitude-limited information where each landing site is summarized by a mean score and a designed ambiguity proxy that contracts as the vehicle descends and senses within a cone-shaped footprint under an altitude-to-resolution schedule. Using this abstraction, we derive closed-form, risk-averse site scoring techniques—an entropic certainty-equivalent and a Gaussian Conditional Value at Risk surrogate—and pair them with greedy and exploratory planners to prioritize sites that are both high-value and robust to late-revealed terrain detail. These rollout-free heuristics improve lower-tail landing outcomes (1st percentile and certainty-equivalent) relative to mean-based baselines, with the largest gains when refinement occurs late and unresolved detail is large. We also demonstrate that these methods perform comparably to or better than Monte Carlo Tree Search baselines with orders-of-magnitude faster computation. Our results are supported by numerical simulations.
\end{abstract}

\section{INTRODUCTION}


In aerospace systems, powered descent and landing requires committing to a touchdown site with significant uncertainty in terrain safety, limited sensing capabilities, decreasing time-to-go, and tight onboard compute for online replanning \cite{YANG2022610,nelson2022landing}. The lander must trade off reaching a high-value (or safe) landing site while respecting the dynamics constraints of descent. These features make descent planning fundamentally different from standard navigation: decisions are irreversible, the opportunity to gather information shrinks rapidly near touchdown, and poor choices can be catastrophic \cite{graydon2020guidance}. Robust descent-planning is especially critical as the volume of autonomous landings is expected to increase with growing interest in unmanned aircraft \cite{ZengEmerg,saldiran2025ensuring}, air taxi services \cite{WEI2022306,graydon2020guidance}, and extraterrestrial landings \cite{scorsoglio2025meta,lee2025challenges}. For human-rated aerospace systems, mission acceptability is often defined by tail-risk metrics such as probability of loss-of-crew rather than average performance alone \cite{NASA_8705_2B}, which makes robust landing decisions especially important in safety-critical descent phases.

Typical state-of-the-art site-selection approaches either select a site from a deterministic map \cite{moghe2020deep,DTMSiteSelection,zha2021landing} or maximize a probabilistic safety metric over candidate sites \cite{tomita2025Mapping,marcus2024landing,tomita2026powered}. However, these methods do not explicitly distinguish between predicted site value and the amount of unresolved information at the current altitude. Separating these concepts lets a planner prefer sites that are not only high-value, but also robust to terrain risk revealed later in descent. For example, Arora et al. account for uncertainty when identifying the best site, but the objective emphasizes confidence of global optimality rather than directly managing lower-tail touchdown outcomes under late hazard revelation during descent~\cite{arora2017multi}. In descent, this can be misaligned with mission objectives: a planner may rationally prefer a slightly lower-valued site with higher reliability, rather than pursue a possibly better site whose value is highly sensitive to information that will only become available late in flight.

\begin{figure}[t]
    \centering
    \includegraphics[width=\linewidth]{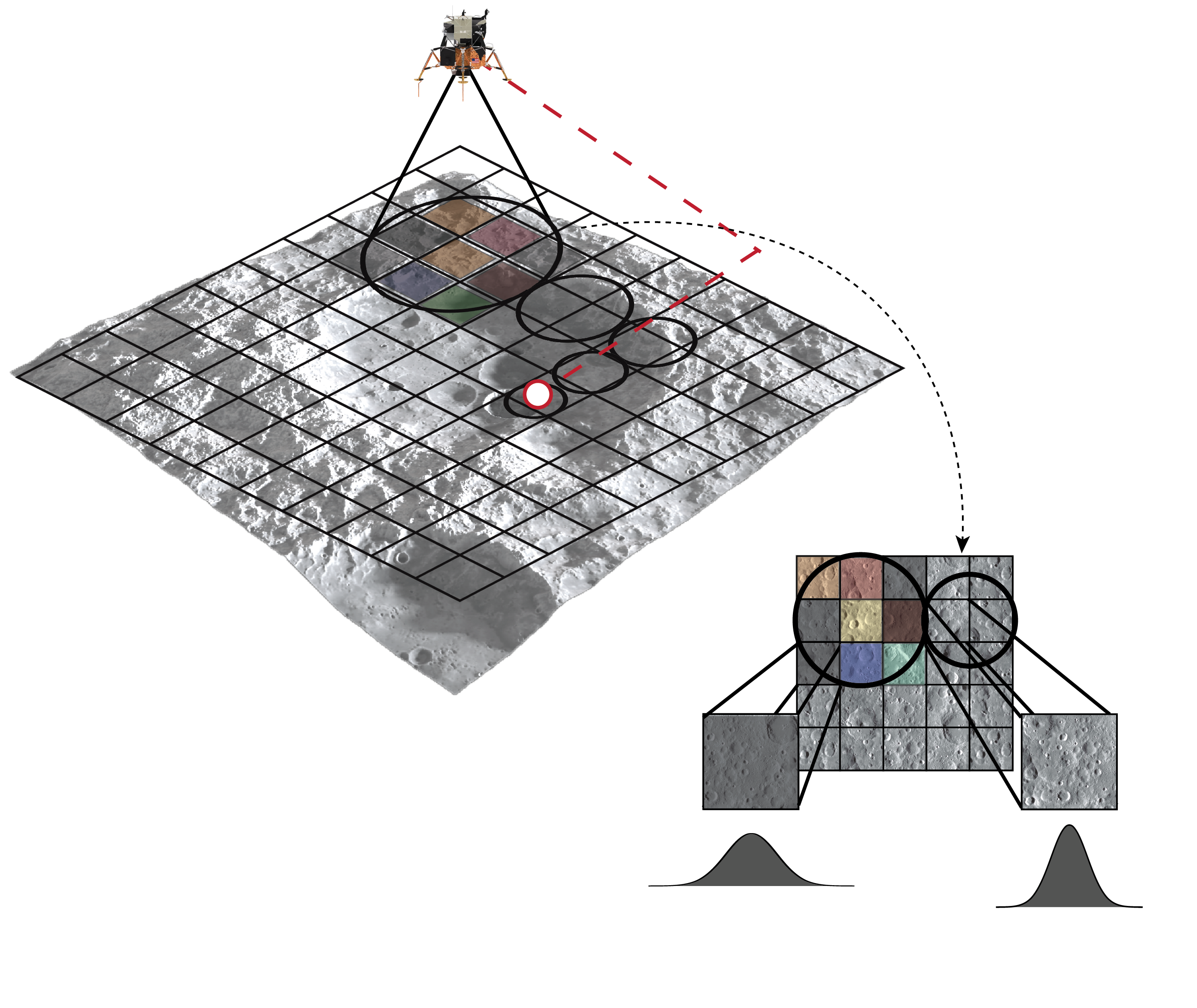}
    \caption{Lunar site-selection problem with altitude-conditioned reachability and progressive terrain refinement over a cone-shaped sensing footprint.}
    \label{fig:domain}
\end{figure}

This paper adopts a deliberately lightweight abstraction of this phenomenon. We first discretize the potential landing area into a fixed set of candidate sites, where each grid cell can be viewed as a point of interest whose score summarizes the surrounding terrain patch. We model the terrain value at each cell as an unknown quantity, and we represent the lander’s evolving knowledge through a mean estimate together with a standard deviation that serves as a designed ambiguity proxy for ``information remaining to be resolved'' about that cell at the current altitude. This proxy is not intended to be a physically exact sensor noise model or a Bayesian posterior; rather, it captures the intuitive fact that certain terrain features (e.g., rocks, craters, slopes) are only resolvable once the vehicle descends below a sensing-dependent altitude threshold. As altitude decreases, a cone-shaped footprint of cells is progressively refined according to a deterministic altitude-to-resolution schedule (\Cref{fig:domain}). In particular, we treat the terrain as fixed but initially unresolved within an episode, making the true value conditioned on the underlying refinement map.

There has been extensive prior work on uncertainty-aware path and trajectory planning in aerospace settings where the final objective site is pre-selected \cite{UncertaintyAwarUAVTraj,RiseAwarePathRover}, but comparatively less work on target/site selection during descent when site value and altitude-limited resolvability are treated as separate quantities. The central question we study is: given altitude-limited information refinement and reachability constraints, how can a lander efficiently select a landing site to avoid poor outcomes?

Accordingly, this work presents low-computation, risk-averse heuristics for site selection and planning during powered descent on a progressively refined grid map. We evaluate closed-form, tail-sensitive cell-scoring rules (certainty equivalent and Gaussian-surrogate CVaR) paired with one-step greedy and exploration-biased planners, and compare them to heuristic and MCTS baselines, and to high-information MCTS references, in deterministic simulations.

\section{PROBLEM FORMULATION}
We model a discrete-time powered descent over a planar grid map with altitude-limited terrain information. The vehicle state is
\begin{equation*}
s_t=(i_t,j_t,z_t)\in \{1,\dots,n_r\}\times\{1,\dots,n_c\}\times\{0,\dots,z_0\},
\end{equation*}
where $(i_t,j_t)$ is the lateral cell index and $z_t$ is a discrete altitude (time-to-go). The action set is
\begin{equation*}
\mathcal{A}=\{\texttt{up},\texttt{down},\texttt{left},\texttt{right},\texttt{none}\}.
\end{equation*}
Each step decreases altitude by one until touchdown ($z=0$), so the horizon is fixed and finite.

\paragraph{Deterministic motion model}
Given $s_t=(i_t,j_t,z_t)$ and action $a_t\in\mathcal{A}$, the next state is deterministic:
\begin{equation*}
s_{t+1}=T(s_t,a_t)=(i_{t+1},j_{t+1},z_{t+1}),
\end{equation*}
with
\begin{equation*}
z_{t+1}=\max(z_t-1,0),
\end{equation*}
and $(i_{t+1},j_{t+1})$ obtained by moving one cell in the chosen direction (or staying for \texttt{none}), saturated at the grid boundaries.
To model fuel/actuation usage, each nontrivial lateral move incurs a fixed step cost
\begin{equation*}
c(a)=
\begin{cases}
0 & \text{if } a=\texttt{none},\\
c_{\text{move}} & \text{if } a\in\{\texttt{up},\texttt{down},\texttt{left},\texttt{right}\},
\end{cases}
\end{equation*}
with $c_{\text{move}}>0$.

\paragraph{Terrain truth}
Let $Y(i,j)$ denote the true terrain value/score at cell $(i,j)$ (e.g., landing quality or safety score). The true terrain information is fixed during descent and decomposed as
\begin{equation*}
Y(i,j)=\mu_{\text{init}}(i,j)+\Delta(i,j),
\end{equation*}
where $\mu_{\text{init}}$ is a coarse prior map available before descent and $\Delta(i,j)$ is an unresolved micro-terrain refinement map (e.g., rocks/slopes) that is initially unobserved and progressively revealed as altitude decreases.
At touchdown ($z_t=0$), the current cell becomes fully known:
\begin{equation*}
\sigma_t(i_t,j_t)\leftarrow 0,\qquad
\mu_t(i_t,j_t)\leftarrow \mu_{\text{init}}(i_t,j_t)+\Delta(i_t,j_t).
\end{equation*}

\paragraph{Agent knowledge as an uncertainty surrogate}
Uncertainty in $Y$ is epistemic (due to altitude-limited resolvability), not aleatoric process noise. The agent maintains a two-parameter uncertainty surrogate
\begin{equation*}
Y(i,j) \sim \mathcal{N}\!\big(\mu_t(i,j),\,\sigma_t(i,j)^2\big),
\end{equation*}
where $\mu_t$ is the current mean estimate and $\sigma_t$ is a designed ambiguity proxy (not a Bayesian posterior) intended to capture “information remaining to be resolved” and enable closed-form, low-compute site scoring. Initially,
\begin{equation*}
\mu_0(i,j)=\mu_{\text{init}}(i,j),\qquad \sigma_0(i,j)=\sigma_{\text{noise}}.
\end{equation*}
In this implementation, the initial grid $\mu_{\text{init}}$ was generated as a smoothly varying field of score means, allowing the planner to anticipate regions that may have higher landing site values based on limited pre-descent information.

\paragraph{Altitude-limited observation model}
At time $t$, before selecting the next action, the lander receives a sensing update over a cone footprint centered at $(i_t,j_t)$ with radius
\begin{equation*}
r(z_t)= z_t \tan(\theta),
\end{equation*}
where $\theta$ is the sensor's cone half-angle. For any cell $c=(i,j)$ within the footprint,
\begin{equation*}
\sqrt{(i-i_t)^2+(j-j_t)^2}\le r(z_t),
\end{equation*}
the belief mean and ambiguity are updated toward the refined value $\mu_{\text{init}}(i,j)+\Delta(i,j)$ with an altitude-dependent learning rate $w(z_t)\in[0,1]$:
\begin{equation*}
\sigma_t(i,j)\leftarrow \min\!\big(\sigma_t(i,j),\ \sigma_{\text{noise}}(1-w(z_t))\big),
\end{equation*}
\begin{equation*}
\mu_t(i,j)\leftarrow \mu_{\text{init}}(i,j) + w(z_t)\,\Delta(i,j).
\end{equation*}
Outside the footprint $(\mu_t,\sigma_t)$ are unchanged.

%

The update is designed to model altitude-limited resolvability rather than a Bayesian posterior update from point measurements. At high altitude, the onboard sensor and reference map resolve only coarse terrain structure, meaning the lander relies primarily on the prior $\mu_{\text{init}}$ and maintains a large ambiguity proxy. As altitude decreases, the sensing resolution improves and fine-scale hazards (captured by the fixed refinement map $\Delta$) become detectable within the cone footprint, so the mean estimate is driven toward the refined value $\mu_{\text{init}}+\Delta$ while the ambiguity proxy contracts to zero in a gradual transition.

To define this gradual transition, represented by the learning weight function $w$, let $z_{\text{upd}}$ denote the altitude at which sensing begins to transition from uninformative to informative (modeling an altitude at which onboard camera/LiDAR data has higher resolution than the onboard reference map). We use $w(z)$ as a designed information-release schedule that approximates increasing resolution with decreasing altitude. The learning weight function is
\begin{equation*}
w(z)=
\begin{cases}
0 & \text{if } z> z_{\text{upd}},\\
1-\dfrac{z}{z_{\text{upd}}} & \text{if } 0\le z\le z_{\text{upd}},\ k=0,\\[8pt]
\dfrac{e^{-k z/z_{\text{upd}}}-e^{-k}}{1-e^{-k}} & \text{if } 0\le z\le z_{\text{upd}},\ k>0
\end{cases}
\end{equation*}
where the learning rate $k\ge 0$ controls the transition sharpness (linear for $k=0$). Higher values of $k$ indicate learning that accelerates closer to the ground. The schedule $w(z)$ is assumed known as part of the altitude-to-resolution model.

Together, these elements define a finite-horizon online site-selection problem: at each altitude, the lander chooses an action from the current belief map, subject to grid motion and altitude-limited sensing, with the goal of reaching a high-value touchdown site while limiting lower-tail risk. The planners below approximate this objective using closed-form scores on the reachable cells.

This model is intended to isolate the planning effect of late information revelation rather than to serve as a calibrated terrain or sensor model. The fixed grid, discrete actions, and Gaussian ambiguity surrogate trade physical fidelity for predictable onboard computation.


\subsection{Risk-Averse Performance Metric}
\label{sec:entropic_metric}

We introduce the exponential transform
\begin{equation*}
U(v) = \exp(-\beta v),\qquad \beta>0
\end{equation*}
which is widely used in risk-sensitive control and decision theory as an exponential utility/penalty \cite{HowardRiskSensitive}. Since $U$ is convex and decreases in $v$, low-performing outcomes (small $v$) incur disproportionately large penalty values.

To summarize performance across a Monte Carlo of $N$ runs, let $v_k\in\mathbb{R}$ denote the scalar outcome of run $k$.

We aggregate penalties across runs and map back to the original value scale via the certainty-equivalent (also known as the entropic risk measure)

\begin{equation*}
\mathrm{CE}(v_{1,\dots,N})
= -\frac{1}{\beta}\log\!\left(\frac{1}{N}\sum_{k=1}^{N} e^{-\beta v_k}\right)
\end{equation*}

This metric is dominated by the lower tail of $\{v\}$: poorer outcomes increase $e^{-\beta v_k}$ sharply, decreasing $\mathrm{CE}$ and thereby penalizing policies with rare but severe failures. This certainty-equivalent therefore acts as a weighted average which emphasizes tail-performance.

\section{TESTED STRATEGIES}

\subsection{Risk-Averse Selection Strategies}
\label{subsec:strats}

\subsubsection{Entropic Tail Selection}
Our evaluation emphasizes lower-tail outcomes via an exponential penalty \Cref{sec:entropic_metric}. Here we use the same entropic form to define a closed-form, risk-averse cell scoring rule for site selection. For a scalar landing outcome $X$, we define the (decreasing) exponential penalty
\begin{equation*}
U(X)=e^{-\beta X},\qquad \beta>0
\end{equation*}
The corresponding certainty-equivalent (entropic risk) maps the penalized expectation back to the original value scale
\begin{equation*}
\mathrm{CE}(X)
:=-\frac{1}{\beta}\log \mathbb{E}\!\left[e^{-\beta X}\right]
\end{equation*}

In our deterministic refinement model, we use $X_t(c)$ as the agent's Gaussian scoring surrogate for the unresolved true terrain value $Y(c)$. Each cell $c$ at time $t$ is summarized by a mean $\mu_t(c)$ and an ambiguity proxy $\sigma_t(c)$:
\begin{equation*}
X_t(c)\sim\mathcal{N}\!\big(\mu_t(c),\,\sigma_t(c)^2\big)
\end{equation*}
This Gaussian form is not derived from stochastic measurements; it is used solely to map $(\mu_t,\sigma_t)$ into tail-sensitive scalar scores.

For $X\sim\mathcal{N}(\mu,\sigma^2)$, the certainty-equivalent admits the closed form
\begin{equation*}
\mathrm{CE}(X)=\mu-\frac{\beta}{2}\sigma^2
\label{eq:entropic-ce}
\end{equation*}
Thus, maximizing $\mathrm{CE}$ trades off high predicted value (large $\mu$) against low ambiguity (small $\sigma$), thereby penalizing sites whose value may change substantially as additional terrain detail is resolved later in descent.

To express this objective as a lower-quantile cell score, we choose a planning quantile that matches the certainty-equivalent. Writing the Gaussian $p$-quantile as
\begin{equation*}
Q_p(X)=\mu+\sigma z_p,\qquad z_p:=\Phi^{-1}(p),
\end{equation*}
where $\Phi$ is the standard normal cumulative distribution function and $z_p$ is the corresponding standard-normal quantile,
the matching condition $Q_p(X)=\mathrm{CE}(X)$ yields an implied z-score
\begin{equation*}
z^\star(\sigma)=-\frac{\beta}{2}\sigma
\end{equation*}
Our implementation uses
\begin{equation*}
z=-\frac{\beta}{2}\sigma_\text{ref}
\end{equation*}
where $\sigma_{\mathrm{ref}}$ is chosen either cellwise ($\sigma_{\mathrm{ref}}=\sigma_t(c)$) or as a summary statistic over the currently reachable set (e.g., mean/max/min standard deviation). These variants were tested through using the simulation. The risk-sensitive cell score is then computed as the Gaussian quantile $Q_p(X)$ and maximized by the planner.

\subsubsection{CVaR}
An alternative tail-focused criterion is the Conditional Value-at-Risk (CVaR), a coherent risk measure that evaluates the expected value of a random variable in its worst $\alpha$–fraction of outcomes.

Using the same Gaussian parameterization $X\sim\mathcal{N}(\mu,\sigma^2)$ as a scoring surrogate for a cell with summary $(\mu_t(c),\sigma_t(c))$, the lower-tail CVaR at level $\alpha\in(0,1)$ is
\begin{equation*}
\mathrm{CVaR}_\alpha(X)
:= \mathbb{E}\!\left[X \,\middle|\, X \leq \mathrm{VaR}_\alpha(X)\right]
\end{equation*}
where $\mathrm{VaR}_\alpha(X)$ is the $\alpha$–quantile of $X$ \cite{tamkin2019distributionally}.
For a Gaussian $X\sim\mathcal{N}(\mu,\sigma^2)$, the lower-tail CVaR admits the closed form
\begin{equation*}
\mathrm{CVaR}_\alpha(X)
= \mu - \sigma\,\frac{\phi(z_\alpha)}{\alpha}
\qquad
z_\alpha := \Phi^{-1}(\alpha)
\label{eq:cvar-normal}
\end{equation*}
where $\phi$ and $\Phi$ are the standard normal pdf and cdf, respectively.

Accordingly, we score each candidate landing cell $c$ at time $t$ by $\mathrm{CVaR}_\alpha\!\big(X_t(c)\big)$ and select actions that maximize this score (after including travel and exploration terms in \Cref{sec:planning}). Smaller $\alpha$ yields more conservative behavior by emphasizing deeper lower-tail outcomes.

We also considered the Entropic Value-at-Risk (EVaR) \cite{ahmadi2012entropic}, defined as
\begin{equation*}
\mathrm{EVaR}_\alpha(L)
:= \inf_{\lambda>0}
\left\{
\frac{1}{\lambda}
\left(
\log \mathbb{E}\!\left[e^{\lambda L}\right]
- \log \alpha
\right)
\right\}
\end{equation*}

Under the Gaussian surrogate used here, both CVaR and EVaR reduce to the same functional form
\begin{equation*}
\mu - k(\alpha)\sigma
\end{equation*}
differing only in how $\alpha$ maps to the penalty magnitude \cite{ahmadi2012entropic}; we therefore report CVaR results as a representative member of this family while computing $\mathrm{CVaR}_\alpha$ in closed form \cite{tamkin2019distributionally}.

\subsection{Planning}\label{sec:planning}
At each time $t$ with state $s_t=(i_t,j_t,z_t)$ and belief $(\mu_t,\sigma_t)$, the planner considers candidate targets within the reachable set
\begin{equation*}
\mathcal{R}_t:=\{(i,j): \|(i,j)-(i_t,j_t)\|_1 \le z_t\}
\end{equation*}
Each candidate cell $c\in\mathcal{R}_t$ is scored using a risk-sensitive value
\begin{equation*}
V_t(c)=\rho\!\big(\mu_t(c),\sigma_t(c)\big)
\end{equation*}
with $\rho(\cdot)$ instantiated as the selected cell-scoring rule, such as the mean score, entropic CE/quantile score, or Gaussian CVaR score. The MCTS variants below are separate search procedures, but use these same cell scores in their rollout policies.

To reflect the per-step movement cost $c_{\text{move}}$, we penalize distant targets using the Manhattan distance
\begin{equation*}
d_t(c):=\|(i_t,j_t)-c\|_1
\end{equation*}
which upper-bounds the number of lateral moves required to reach $c$ (ignoring boundary saturation).

\subsubsection{Greedy Actions}
The greedy planner selects a target cell by
\begin{equation*}
c_t^\star \in \arg\max_{c\in\mathcal{R}_t}\Big(V_t(c)-\lambda_{\text{travel}}\,d_t(c)\Big)
\end{equation*}
and then executes the one-step action that minimizes the straight-line distance to $c_t^\star$.

\subsubsection{Exploratory Actions}
\label{sec:explorationPlanning}
To encourage information gathering, we add an action-dependent exploration bonus. For each action $a\in\mathcal{A}$, let $s_{t+1}(a)=T(s_t,a)$ be the next state and define a heuristic gain
\begin{equation*}
G_t(a)=\!\!\sum_{c\in \mathcal{F}(s_{t+1}(a))}
\max\!\big(0,\sigma_t(c)-\sigma_{\text{new}}(z_{t+1})\big)\,\mu_t(c)
\end{equation*}
where $\mathcal{F}(\cdot)$ is the cone footprint and $\sigma_{\text{new}}(z)=\sigma_{\text{noise}}(1-w(z))$ is the post-update standard deviation implied by the sensing model The multiplicative factor $\mu_t(c)$ biases exploration toward refining cells that are currently predicted to be valuable, rather than spending sensing effort on uniformly low-value regions. For a target $c$, let $a(c)$ denote the one-step move toward $c$. The target score becomes
\begin{equation*}
S_t(c)=V_t(c)-\lambda_{\text{travel}}\,d_t(c)+\lambda_{\text{explore}}\,G_t\!\big(a(c)\big)
\end{equation*}
and the selected target is $c_t^\star\in\arg\max_{c\in\mathcal{R}_t} S_t(c)$. Setting $\lambda_{\text{explore}}=0$ recovers the greedy rule above.

\subsection{Baselines}
\label{subsec:baselines}

\subsubsection{Monte Carlo Tree Search}
\label{sec:MCTS}

We compare the introduced risk-averse selection and planning methods against five variants of Monte Carlo Tree Search (MCTS). MCTS is a powerful anytime algorithm for online planning under uncertainty, iterating over four steps: selection, expansion, simulation, and backpropagation. In this setting, tree nodes represent simulated lander belief states, edges represent actions in $\mathcal{A}$, and depth corresponds to descent time.

\textbf{Selection:}
Beginning at the root state node $s_0$, the tree is traversed by repeatedly selecting actions and successor state nodes until reaching a leaf. This is often accomplished using Upper Confidence Bound for Trees (UCT) exploration
\begin{align*}
    a \leftarrow \arg\max_{a} \left[ \hat{Q}(s, a) + c\sqrt{\frac{\ln{N(s)}}{N(s, a)}} \right]
\end{align*}
where $N(s)$ and $N(s,a)$ denote state and state-action visit counts, respectively, $c$ is the exploration constant, and $\hat{Q}(s,a)$ is the estimated action value.

\textbf{Expansion:}
Generate child state node $s_L$ on arriving at a non-terminal leaf node. Initialize visit counts and action-values to zero.

\textbf{Simulation:}
Execute a rollout policy, random or otherwise, to a terminal state or some defined horizon. Maintain the return $G$. We employ a greedy rollout policy for all MCTS variants. 

\textbf{Backpropagation:}
The simulation return $G$ is propagated back along the traversed path with incremental updates: $N(s, a) \leftarrow N(s, a) + 1$ and $\hat{Q}(s, a) \leftarrow \hat{Q}(s, a) + (G - \hat{Q}(s, a))/N(s, a)$.

\paragraph{High-information references}
We introduce two high-information MCTS references for comparison. First, an oracle MCTS has perfect knowledge of the true terrain values. Second, a blackbox MCTS has access to the true observation function. Blackbox MCTS approaches oracle MCTS in the limit of sufficient search, but with a finite rollout budget its performance can vary across metrics and may be slightly exceeded by a heuristic in some regimes. Such cases reflect incomplete convergence and finite-sample variability rather than a strict violation of a mathematical upper bound.

\paragraph{Risk-Averse}
We also consider three variants of MCTS that assume no knowledge of the observation function, relying instead on CVaR, max $\sigma$ entropic tail selection, or mean statistics generated using the most updated belief map. 

\begin{table}[ht]
\centering
\caption{MCTS Hyperparameters}
\label{tab:mcts_hyperparams}
\begin{tabular}{@{}lr@{}}
\toprule
\textbf{Parameter} & \textbf{Value} \\
\midrule
UCT iterations per action & 700 \\
Exploration constant ($c$) & 3.0 \\
Rollout policy & Greedy on cell score \\
Cell scoring function & $\{$Mean, CVaR$_\alpha$, Entropic-$\sigma$$\}$ \\
Action selection at root & Max mean Q-value \\
Action space & $|\mathcal{A}| = 5$ \\
\bottomrule
\end{tabular}
\end{table}

\subsubsection{Mean Targeting}

A greedy mean targeting strategy was also tested which ignores the remaining knowledge for each cell ($\sigma_t(c)$) and instead attempts to maximize the performance based on the prior and observed information. This strategy was tested with and without the exploration planning mention in \Cref{sec:explorationPlanning}.

\section{RESULTS}

We evaluate whether closed-form, risk-averse cell scoring can improve lower-tail touchdown outcomes under late terrain refinement while remaining computationally lightweight. The main comparisons are between mean and risk-sensitive cell scores, greedy and exploratory action selection, and lightweight heuristics versus high-information MCTS references. We report mean landing score, first percentile (P1), and certainty-equivalent (CE) across Monte Carlo trials, and compare against mean-targeting baselines and high-information MCTS references. The aforementioned simulations were run with the discussed site-selection and planning strategies along with the benchmarks. The entropy penalty parameter (and entropic tail selection parameter) $\beta$ was chosen as $0.5$. The results show the performance for CVaR parameter $\alpha = 0.80$ for the greedy planning case and $\alpha = 0.25$ for the exploratory planning case, as explained in this section. For the entropic tail selection, both a cell-wise approach and a maximum $\sigma$ approach were utilized. The initial grid was generated to range from $-10$ to $10$. The sensor's half cone angle was chosen as $\theta = \pi/8$ and the agent's travel penalty was chosen as $c_{\text{move}}=0.01$.

\begin{figure}
    \centering
    \includegraphics[width=\linewidth]{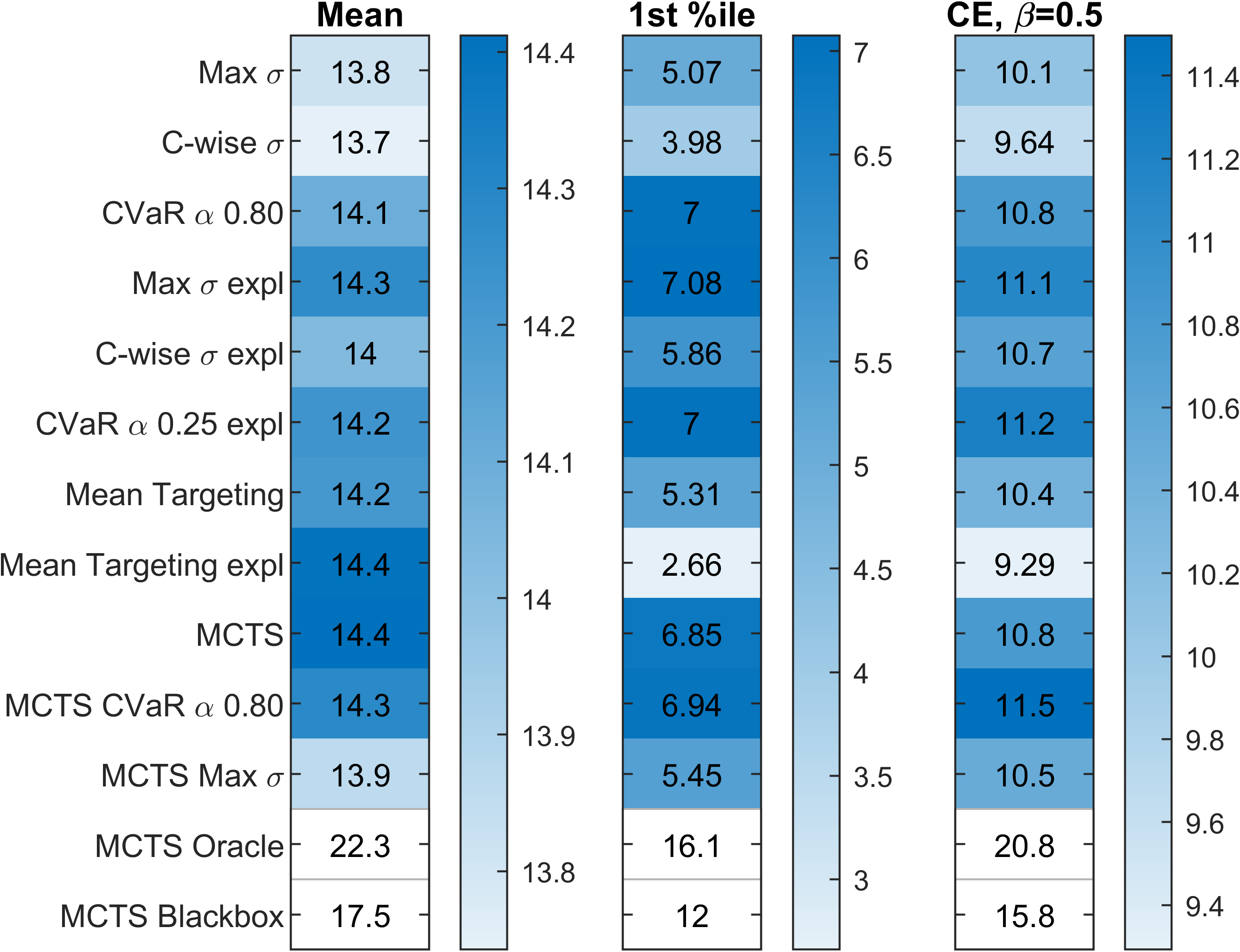}
    \caption{Mean landing score, first percentile landing value, and certainty-equivalent for the simulations with $k = 15, \sigma_\text{noise} = 5$. MCTS Oracle and MCTS Blackbox serve as high-information references only, as mentioned in \Cref{sec:MCTS}.}
    \label{fig:primary_results}
\end{figure}

Of the tested values of the learning rate $k$ ($1,2,4,7,15$) and global knowledge gap noise $\sigma_\text{noise}$ ($0.5, 1, 2, 3, 5$), the trends between cases stayed consistent, and the performance differences were most exaggerated in the extreme $k = 15, \sigma_\text{noise} = 5$ case. For this test sweep, simulations were run over 1000 landing grids, each with identical layouts of the initial belief and information to gain maps across the different site-selection and planning strategies. The results for this case are shown in \Cref{fig:primary_results}.

These results show that although a simple mean targeting strategy fares well when analyzing the average landing value across all runs with the risk-averse methods not far behind in performance, its tail-end performance does not fare as well. For both the first percentile and the certainty-equivalent, each of the exploratory planning methods applied to the risk-averse strategies, along with CVaR using the best-performing $\alpha$, performed better than the mean targeting method. It should be noted that on the tail end, exploration actually hinders the performance of mean targeting.

The strongest non-MCTS strategies in this regime are max-$\sigma$ entropic scoring with exploratory planning and tuned CVaR with exploratory planning. Mean targeting remains competitive on average landing score, but it is weaker in the lower tail.

Among the entropic variants, the max-$\sigma$ version consistently outperforms the cell-wise version. Under altitude-limited refinement, uncertainty becomes highly nonuniform as some cells appear low-$\sigma$ simply because the footprint has partially refined them earlier. Using $\sigma_{\mathrm{ref}}=\max_{c\in\mathcal{R}_t}\sigma_t(c)$ maintains conservatism until refinement has progressed broadly enough, reducing a late-discovery failure mode and improving tail outcomes.

The MCTS variants provide compute-heavy reference performance by explicitly searching over future refinement and reachability. In contrast, our heuristics use only closed-form cell scores and one-step planning. While MCTS typically attains stronger performance than simple baselines, it requires hundreds of rollouts per action (\Cref{tab:mcts_hyperparams}), whereas the proposed heuristics are designed for real-time onboard evaluation over large candidate sets. However, only with the certainty equivalent metric of the optimal CVaR strategy does the MCTS implementation outperform the simple exploratory planning when applied to their respective site-selection strategies.

\begin{table}
\centering
\caption{Mean and Median compute time per step for the risk-averse strategies and MCTS baselines.}
\label{tab:compute_time}
\begin{tabular}{@{}lrr@{}}
\toprule
Policy & Mean (ms) & Median (ms) \\
\midrule
\textbf{Risk aware:} & &  \\
Max $\sigma$ & 1.38 & 0.48 \\
CVaR $\alpha \; 0.80$ & 0.06 & 0.03 \\
Max $\sigma$ expl & 0.95 & 0.41 \\
CVaR $\alpha \; 0.25$ expl & 0.04 & 0.03 \\
\midrule
\textbf{MCTS:} & & \\
Mean & 30.69 & 11.74 \\
CVaR $\alpha \; 0.80$ & 31.41 & 13.15 \\
Max $\sigma$ & 30.94 & 13.53 \\
Oracle & 32.69 & 13.37 \\
Blackbox & 31.59 & 13.59 \\
\bottomrule
\end{tabular}
\end{table}

Crucially, the proposed methods avoid online rollouts, requiring only per-cell algebraic scoring and one-step action selection, making them suitable when onboard compute or time-to-go preclude deep search. Compute times are shown in \Cref{tab:compute_time}, measured single-threaded on an AMD Ryzen 9 9900X3D. The max-$\sigma$ entropic strategy is slowed mainly by reachable-set computation, which is often part of standard lander site-selection pipelines \cite{tomita2026powered}. Even so, entropic tail selection is 1 to 2 orders of magnitude faster than MCTS, and CVaR is three orders faster. For landers evaluating hundreds or thousands of candidate sites under varying illumination, sensor geometry, and sensor limits such as LiDAR periphery distortion, these closed-form scores substantially reduce onboard computation.

When reflecting on the gap to these high-information references, it is clear that the proposed heuristics do not reach the true optimal solution, especially in the first-percentile tail metric. This is expected because $k=15$ delays refinement until late in descent and $\sigma_{\text{noise}}=5$ increases the magnitude of unresolved detail, amplifying the late-discovery failure mode that dominates lower-tail outcomes. \Cref{tab:gap_to_upperbounds} reports, for each $(k,\sigma_{\text{noise}})$ regime, the gaps to two high-information references (MCTS Oracle and MCTS Blackbox) across Mean, first percentile (P1), and the entropic score (CE). In each regime, we define \emph{Best} as the non-reference heuristic with the highest CE, and we report that strategy’s gaps on all three metrics. Note that the blackbox entries should be interpreted as finite-compute references; occasional negative gaps arise from finite rollout budgets and Monte Carlo variability. Two consistent trends emerge. First, for fixed $\sigma_{\text{noise}}$, the gap increases with larger transition parameter $k$ (e.g., $k=15$ exhibits larger gaps than $k=0$), indicating that the extreme transition regime is substantially harder and leads to greater suboptimality relative to the high-information references. Second, for fixed $k$, the gap increases as $\sigma_{\text{noise}}$ grows from $0.5$ to $5.0$, with the largest discrepancies occurring at $(k,\sigma_{\text{noise}})=(15,5.0)$. This effect is most pronounced in the tail metric P1, where the gaps to the oracle high-information reference become largest, highlighting that conservative lower-tail performance is particularly difficult to match in high-noise, high-transition settings.

\begin{table}
\centering
\caption{Gap between the CE-selected best heuristic and two high-information references. For each $(k,\sigma_{\text{noise}})$ regime, the risk-averse strategy with the highest certainty-equivalent (CE) is used. Entries report $(\text{Reference}-\text{Best})$ for Mean, P1, and CE; smaller is closer to the high-information reference.}
\label{tab:gap_to_upperbounds}
\setlength{\tabcolsep}{2.5pt}
\renewcommand{\arraystretch}{1.1}
\begin{tabular}{@{}r r p{1.8cm} rrr rrr@{}}
\toprule
$k$ & $\sigma_{\text{noise}}$ & Best (by CE) &
\multicolumn{3}{c}{MCTS Oracle} & \multicolumn{3}{c}{MCTS Blackbox} \\
\cmidrule{4-6}\cmidrule{7-9}
 &  &  & Mean & P1 & CE & Mean & P1 & CE \\
\midrule
0  & 0.5 & Max $\sigma$ Expl              & 0.181 & 0.398 & 0.196 & 0.059 & 0.276 & 0.067 \\
4  & 0.5 & Max $\sigma$ Expl              & 0.230 & 0.514 & 0.253 & 0.105  & 0.372 & 0.120 \\
15 & 0.5 & Max $\sigma$ Expl              & 0.312 & 0.676 & 0.348 & 0.177  & 0.516 & 0.202 \\
\midrule
0  & 2.0 & C-wise $\sigma$ Expl            & 1.36  & 1.98  & 1.50  & 0.334  & 0.698 & 0.392 \\
4  & 2.0 & CVaR $\alpha\;0.25$ Expl        & 1.75  & 2.66  & 2.00  & 0.637  & 1.15  & 0.778 \\
15 & 2.0 & C-wise $\sigma$ Expl            & 2.66  & 5.00  & 3.22  & 1.38   & 3.16  & 1.79 \\
\midrule
0  & 5.0 & Max $\sigma$                    & 3.31  & 2.44  & 3.30  & 0.053 & -0.317 & 0.033 \\
4  & 5.0 & C-wise $\sigma$ Expl            & 5.19  & 5.01  & 5.66  & 1.16   & 1.79  & 1.51 \\
15 & 5.0 & CVaR $\alpha\;0.25$ Expl        & 8.02  & 9.10  & 9.55  & 3.27   & 4.96  & 4.58 \\
\bottomrule
\end{tabular}
\end{table}

\begin{figure}
    \centering
    \includegraphics[width=\linewidth]{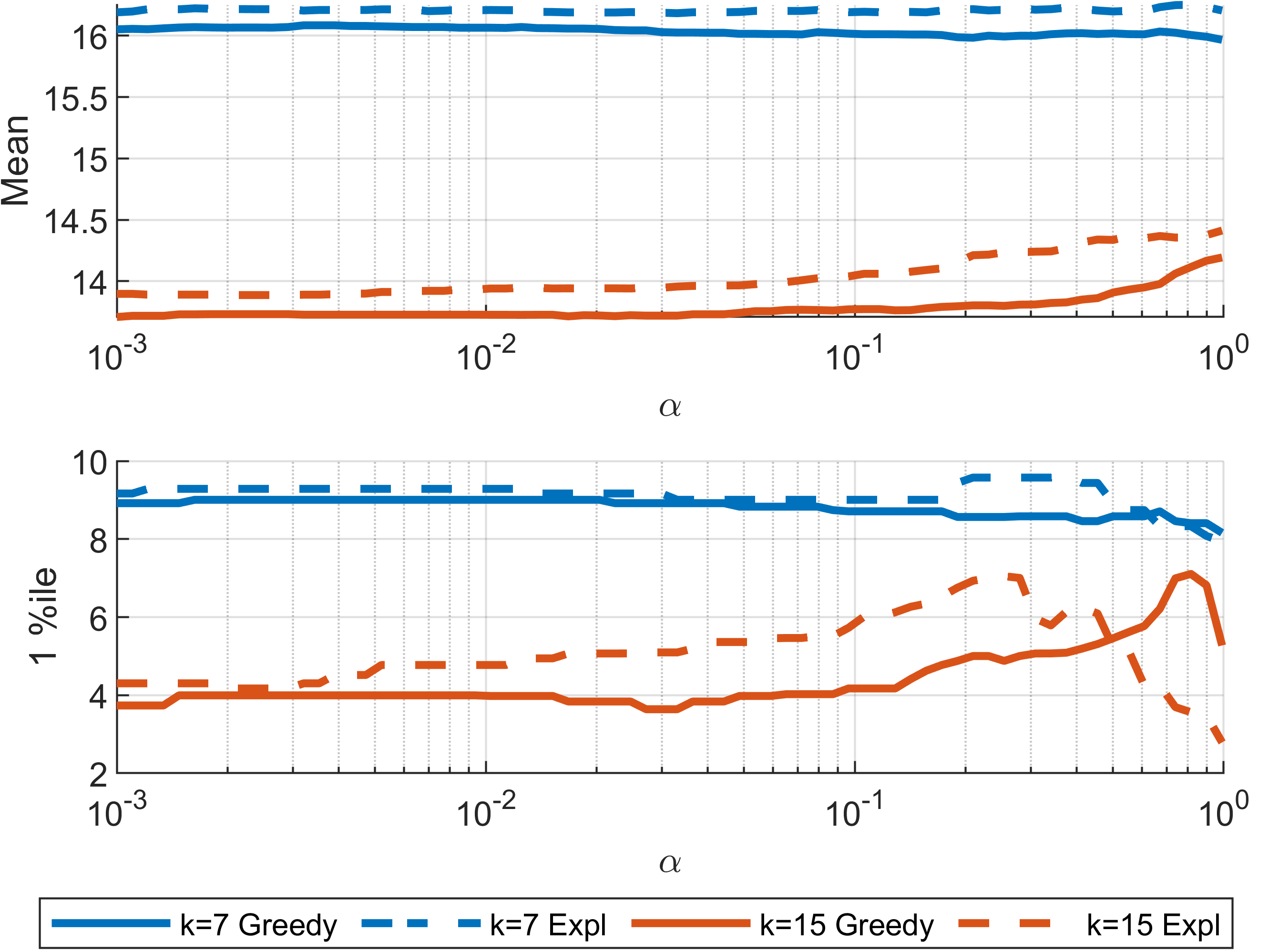}
    \caption{CVaR level sweep at $\sigma_{\text{noise}}=5$ comparing refinement timing ($k=7,15$) and planning mode (greedy, exploratory) across various values of the hyperparameter $\alpha$, Top: mean landing value. Bottom: 1st-percentile landing value.}
    \label{fig:CVaRAblation}
\end{figure}

CVaR is sensitive to $\alpha$. \Cref{fig:CVaRAblation} shows a CVaR sweep at $\sigma_{\text{noise}}=5$ for two refinement-timing regimes ($k=7,15$), reporting mean landing value and first-percentile performance. The best tail-performing $\alpha$ is regime-dependent, and this sweep informed the $\alpha$ values used in the primary results. Values near-optimal for $k=7$ do not remain optimal for $k=15$; in the late-learning case, the lower tail varies substantially with $\alpha$, especially under exploratory planning. This is a practical limitation of fixed-parameter CVaR scoring: selecting $\alpha$ sets a conservatism level whose appropriate value depends on the sensing/refinement regime. In contrast, policies that perform well across a broader range of $\alpha$ are less sensitive to risk calibration. We also observe that for $k=15$, exploration improves the lower tail for intermediate $\alpha$ but degrades it for very large $\alpha$, suggesting that optimism and information-seeking must be balanced under late refinement.

Entropic tail selection showed less sensitivity to its tuning parameter in our experiments. Although it also introduces a hyperparameter through the chosen tail quantile, this parameter primarily sets the conservatism of the lower-tail score rather than a regime-specific risk level. As a result, its performance was more stable across regimes than CVaR, whose best-performing $\alpha$ varied more noticeably with the learning schedule.

\section{CONCLUSION}
We studied landing-site selection under altitude-limited information, where terrain detail is revealed progressively during descent and late refinement can induce severe tail outcomes. Using a Gaussian surrogate over per-cell mean and ambiguity, we derived very efficient, closed-form, and risk-averse scoring rules (entropic tail selection and CVaR) and paired them with simple greedy and exploratory planners. In deterministic Monte Carlo experiments, risk-averse scoring improves lower-tail metrics relative to mean targeting, with the largest benefits in late-learning/high-uncertainty regimes; we also quantified the remaining gap to oracle and blackbox high-information MCTS references. Future work will incorporate stochastic sensing while preserving the primary design goal: efficient, tail-aware site selection suitable for compute- and time-limited onboard descent guidance.

\addtolength{\textheight}{-0cm}   


\section*{ACKNOWLEDGMENT}

The work of V. Patel was supported by the National Science Foundation Graduate Research Fellowship under Grant No. DGE-2146755.


\bibliographystyle{IEEEtran}
\bibliography{myreferences}

@ARTICLE{UncertaintyAwarUAVTraj,
  author={Chang, Mai and Zhou, Jianshan and Tian, Daxin and Duan, Xuting and Qu, Kaige and Cao, Dongpu},
  journal={IEEE Internet of Things Journal}, 
  title={Uncertainty-Aware Robust UAV Trajectory Planning With Dynamic Collision Avoidance}, 
  year={2025},
  volume={12},
  number={17},
  pages={36502-36516},
  doi={10.1109/JIOT.2025.3582721}}

@INPROCEEDINGS{RiseAwarePathRover,
  author={Sánchez-Ibáñez, J. Ricardo and Sanchez-Cuevas, Pedro J. and Olivares-Mendez, Miguel},
  booktitle={IEEE/RSJ International Conference on Intelligent Robots and Systems}, 
  title={Optimal and Risk-Aware Path Planning considering Localization Uncertainty for Space Exploration Rovers}, 
  year={2022},
  volume={},
  number={},
  pages={4092-4099},
  doi={10.1109/IROS47612.2022.9981179}}

@ARTICLE{DTMSiteSelection,
  author={Jung, Youeyun and Lee, Seongheon and Bang, Hyochoong},
  journal={IEEE Transactions on Aerospace and Electronic Systems}, 
  title={Digital Terrain Map Based Safe Landing Site Selection for Planetary Landing}, 
  year={2020},
  volume={56},
  number={1},
  pages={368-380},
  doi={10.1109/TAES.2019.2913600}}

@article{moghe2020deep,
  title={A deep learning approach to hazard detection for autonomous lunar landing},
  author={Moghe, Rahul and Zanetti, Renato},
  journal={The Journal of the Astronautical Sciences},
  volume={67},
  number={4},
  pages={1811--1830},
  year={2020},
  publisher={Springer}
}

@inproceedings{zha2021landing,
  title={Landing site selection of detector based on 3D point cloud segmentation},
  author={Zha, Keke and Yuan, Jiabin},
  booktitle={International Conference on Mechanical, Aerospace and Automotive Engineering},
  pages={104--112},
  year={2021}
}

@article{tomita2025Mapping,
  title={Real-Time Stochastic Terrain Mapping and Processing for Autonomous Safe Landing},
  author={Tomita, Kento and Ho, Koki},
  journal={Journal of Spacecraft and Rockets},
  volume={62},
  number={5},
  pages={1848--1868},
  year={2025},
  publisher={American Institute of Aeronautics and Astronautics}
}

@article{marcus2024landing,
  title={Landing Site Mapping and Selection with Quadtree Map Using Planar Elements},
  author={Marcus, Corey L and Zanetti, Renato and Setterfield, Timothy P},
  journal={Journal of Guidance, Control, and Dynamics},
  volume={47},
  number={7},
  pages={1283--1297},
  year={2024},
  publisher={American Institute of Aeronautics and Astronautics}
}

@inproceedings{tomita2026powered,
  title={Powered Descent Decision Making: A Reachability-Steering Approach},
  author={Tomita, Kento and Elango, Purnanand and Vinod, Abraham P and Di Cairano, Stefano and Weiss, Avishai},
  booktitle={AIAA SciTech Forum},
  pages={0328},
  year={2026}
}

@article{YANG2022610,
title = {Autonomous UAVs landing site selection from point cloud in unknown environments},
journal = {ISA Transactions},
volume = {130},
pages = {610-628},
year = {2022},
issn = {0019-0578},
doi = {https://doi.org/10.1016/j.isatra.2022.04.005},
author = {Linjie Yang and Chenglong Wang and Luping Wang}
}

@article{nelson2022landing,
  title={Landing Site Selection Using a Geometrically Conforming Footprint on Hazardous Small Bodies},
  author={Nelson, Joshua D and Schaub, Hanspeter},
  journal={Journal of Spacecraft and Rockets},
  volume={59},
  number={3},
  pages={889--899},
  year={2022},
  publisher={American Institute of Aeronautics and Astronautics}
}

@inproceedings{graydon2020guidance,
  title={Guidance for designing safety into urban air mobility: Hazard analysis techniques},
  author={Graydon, Mallory and Neogi, Natasha A and Wasson, Kimberly},
  booktitle={AIAA SciTech Forum},
  pages={2099},
  year={2020}
}

@article{WEI2022306,
title = {Safety Verification for Urban Air Mobility Scheduling},
journal = {IFAC Conference on Networked Systems},
volume = {55},
number = {13},
pages = {306-311},
year = {2022},
issn = {2405-8963},
doi = {https://doi.org/10.1016/j.ifacol.2022.07.277},
author = {Qinshuang Wei and Gustav Nilsson and Samuel Coogan},
}

@Article{ZengEmerg,
AUTHOR = {Zeng, Xiwei and Chu, Xinlan and Yao, Rundong and Liu, Jiayi and Yang, Yuezhou and Zeng, Weili},
TITLE = {An Emergency Rescue Reconnaissance UAV Nest Site Selection Method Considering Regional Differentiated Coverage and Rescue Satisfaction},
JOURNAL = {Aerospace},
VOLUME = {12},
YEAR = {2025},
NUMBER = {9},
ARTICLE-NUMBER = {798},
ISSN = {2226-4310},
DOI = {10.3390/aerospace12090798}
}

@article{saldiran2025ensuring,
  title={Ensuring operation time safety of vtol uav: Autonomous emergency landings in unknown terrain},
  author={Saldiran, Emre and Hasanzade, Mehmet and Cetin, Aykut and Inalhan, Gokhan},
  journal={IEEE Transactions on Aerospace and Electronic Systems},
  year={2025},
  publisher={IEEE}
}

@article{scorsoglio2025meta,
  title={Meta-reinforcement learning guidance, navigation, and control for autonomous lunar landing with safe site selection},
  author={Scorsoglio, Andrea and Gaudet, Brian and Ghilardi, Luca and Furfaro, Roberto},
  journal={Neural Computing and Applications},
  volume={37},
  number={22},
  pages={17311--17340},
  year={2025},
  publisher={Springer}
}

@article{lee2025challenges,
  title={The Challenges of Aerospace Engineering in Designing Mars-Landing Systems},
  author={Lee, James},
  journal={American Journal of Aerospace and Aeronautical Engineering},
  volume={6},
  number={5},
  pages={11--15},
  year={2025}
}

@inproceedings{tamkin2019distributionally,
  title={Distributionally-aware exploration for cvar bandits},
  author={Tamkin, Alex and Keramati, Ramtin and Dann, Christoph and Brunskill, Emma},
  booktitle={NeurIPS 2019 Workshop on Safety and Robustness on Decision Making},
  year={2019}
}

@article{ahmadi2012entropic,
  title={Entropic value-at-risk: A new coherent risk measure},
  author={Ahmadi-Javid, Amir},
  journal={Journal of Optimization Theory and Applications},
  volume={155},
  number={3},
  pages={1105--1123},
  year={2012},
  publisher={Springer}
}

@inproceedings{arora2017multi,
  title={Multi-modal Active Perception for Autonomously Selecting Landing Sites on Icy Moons},
  author={Arora, Akash and Furlong, Michael and Wong, Uland and Sukkarieh, Salah and Fong, Terry W},
  booktitle={AIAA SPACE and Astronautics Forum and Exposition},
  pages={5182},
  year={2017}
}

@misc{NASA_8705_2B,
  author       = {{NASA}},
  title        = {Human-Rating Requirements for Space Systems},
  howpublished = {NASA Procedural Requirements 8705.2B},
  year         = {2008},
  month        = may,
}

@article{HowardRiskSensitive,
 ISSN = {00251909, 15265501},
 author = {Ronald A. Howard and James E. Matheson},
 journal = {Management Science},
 number = {7},
 pages = {356--369},
 publisher = {INFORMS},
 title = {Risk-Sensitive Markov Decision Processes},
 urldate = {2026-03-30},
 volume = {18},
 year = {1972}
}

\end{document}